# Business processes integration and performance indicators in a PLM

A. Bissay, P. Pernelle, A. Lefebvre, A. Bouras,
*LIESP – Université de Lyon – France*
aurelie.bissay@marmillon.fr, *[philippe.pernelle][ arnaud.lefebvre]@iutb.univ-lyon1.fr,* abdelaziz.bouras@univ-lyon2.fr,

***Abstract:*** *In an economic environment more and more competitive, the effective management of information and knowledge is a strategic issue for industrial enterprises. In the global marketplace, companies must use reactive strategies and reduce their products development cycle. In this context, the PLM (Product Lifecycle Management) is considered as a key component of the information system. The aim of this paper is to present an approach to integrate Business Processes in a PLM system. This approach is implemented in automotive sector with second-tier subcontractor*

# 1. INTRODUCTION

The reduction of new products development delays leans on the optimization of the development process upstream stages. Hence, information systems managing product development must be optimized.

In many manufacturing companies, the information system lean on a PLM approach. In this context, data management is covered by the PLM system (Sudarsan and al 2005), but process management is more problematic especially in the initial step of product development.

The aim of this paper is to propose an approach to integrate business processes within PLM systems in order to optimize the product development delays as well as the performance.

The first part of this article describes the context of business processes and their assessment through performance indicators. The second part presents an application in the scope of business processes integration.

# 2. MANAGEMENT AND EVALUATION OF THE BUSINESS PROCESSES

## 2.1. Modelling of the Business Processes

The process management within PLM systems encounters some difficulties in many industrial sectors. It's due to the presence of several processes types:
- processes related to product design;
- processes related to change management and data monitoring within the information system;
- processes related to various business;
- collaborative processes.

Theses difficulties are caused by the diversity of processes (in terms of structure and complexity) and their interconnections within the information system. There are several formalisms (MOP, IDEF0, IDEF3, Workflow, etc.) allowing to model and to execute the process (Morley & al. 2005). In this context, the BPM (Business Process Management) approach (Arkin 2002) gives a "bottom-up" overview of all business processes in order to automate and optimize them.

In theory, BPM deals with the whole product lifecycle in spite of the difficulty to model certain steps. Furthermore, the process lifecycle in a BPM approach is the same that those managed by the PLM: Modelling – Instantiation - Executing - Monitoring – Optimization.

Thus, the standardization of the models representation (and models execution) is a major challenge to facilitate integration between the components of the information system. Several approaches exist such as the OMG around BPMN (Business Process Modelling Notation) (OMG 2006) and BPEL (Business Process Execution Language) based on UML and XML. Another approach is the WfMC around XPDL (WFMC 2005).

## 2.2. Performance indicators

The performance has a great interest in the enterprise. However, it remains difficult to evaluate because it depends on many criteria (Capacity to satisfy its customers, optimization of the production system …). The performance indicators are used to characterize this performance. In the PLM systems, the need of indicators must respond to the product development objectives.

As example of multi-criteria indicators, Kaplan and Norton (Kaplan & Norton 2003) present a model of performance measurement, the balanced scorecard, which takes into account several criteria:
- *Financial Perspective*: The financial informers, oriented measures profitability. For instance, the ROI allows evaluating action performance engaged by the past.
- *Customer Perspective*: The indicators of this axis are generally used to evaluate the satisfaction and the fidelity of the customers, the measure of the increase of the clientele and the increase of the profitability by customer …
- *Internal Process Perspective:* This category includes all processes closely contributing to the value creation without omitting the process related to innovation.
- *Learning Perspective:* This axis is used to measure the staff training to access new skills, the information system improvement and the adequation between procedures and practices.

In order to evaluate the performance and adaptability of business processes, we want to construct indicators with PLM tools (Traceability, Workflow, Lifecycle…). The construction of indicators leans on various works concerning the indicators of performances (Berrah 2002). The evaluation of the product development process can be made from qualitative, quantitative, or financial indicators. They can take the form of budget variance, ratios, graphics, flashing, a simple sentence comment, and so on. There are many types of performance indicators such as reporting and piloting indicators, simple and complex, strategic, tactical, operational, and so on. There are different models for implementing indicators, for instance the fuzzy logic (Berrah and al 2006).
In our analysis, there are two types of indicators:
- Performance indicators that are established from the past actions and which are the state of a situation over.
- Process indicators, which conversely, estimate real-time situation. They are used to detect the drifts and to develop corrective action plans.

## 3. AN APPROACH FOR BUSINESS PROCESSES INTEGRATION

The business processes integration in a PLM system (Pernelle P. & al 2006) must respond to various constraints: explicit links with the lifecycles and indicators.
To construct an instanciable model, it is necessary to define the lifecycle of data flow. Indeed, PLM systems lean on a product/lifecycle centric approach. The lifecycle indicates the whole set of phases which could be recognized as independent steps that a product (or data product) might follow. With lifecycles, the process models can not be made independently of data models.
In order to integrate business processes and evaluate them within a PLM system, we propose a typology of processes with associated indicators. The objective is to evaluate them at the time of their execution by the PLM system.

There are different types allowing to caracterize the process: based on activities, on the types of inputs / outputs (Shelton 1996), or on the degree of structuring (Mhamedi and al. 1995).

The typology of the process proposed in the PLM systems, is based on a decomposition following four criteria:
- The time: limited process, not limited, cyclical.
- The stability: stable process, evolutionary, unstable.
- The genericity: single instance, multiple instances.
- The measurability: measurable process, not measurable.

This typology is used in the processes definition. Indeed, the choice of a type imposes some constraints on the various constituents (eg: a process not limited in time will have a state-diagram different from a limited one). Once the model established, process can be executed. Depending on the process type, several instances of a single model can be run simultaneously with a different data context. Moreover, the process type impose constraints on indicators and objectives.

The analysis of the objectives and indicators seems appropriate to characterize the process activities.The objectives are intended to represent the expectations and indicators are tools to check if the results are in line with expectations. In many cases, the objective is not the only elements flow out of the process. In fact, some objectives are difficult to measure. The type of objective is a notion allowing distinguishing the degree of measurability and the continuity of an objective. Indeed, if it is simple to estimate an objective based on a value of threshold or on a production of a technical document, it is not of there even when the end of the activity participates in piloting or in managing the produced information system. For example, a regulatory activity of a working group is not easily measurable in quantitative terms. The continuity distinguishes objectives whose reach is spread over time.
On the one hand, we have objectives which can be reached or not reached but which, once in one of these two states, cannot change it any more.
On the other hand, we have objectives for which there is a transition allowing change from a reached state to not reached state. The basic objectives, as their name suggests, can describe an aim unit (mesureable or not mesureable) not divisied. Starting from these two basic objectives, it is possible to define subcategories adapted to specific functions.

## 4. AN INDUSTRIAL CASE STUDY

### 4.1. Context description

The industrial context of this paper lies in the industrial sector in the plastics industry.
The Marmillon company is a SME, specializes in processing plastic. It works primarily like second-tier subcontractor to the automotive industry. Like many enterprises, the company hired a quality assurance type of ISO 9001 and ISO TS 16949. The processes and quality manuals present an important source of knowledge facilitating the formalization of knowledge and construction business process (Figure2).

### 4.2. Management of request for quotation

The Business Activity in Marmillon company is based on responses to Request For Quotation. We have therefore chosen to focus our analysis on this business process. Furthermore, the positive rate of acceptance is not in accordance with the strategic requirements.

The analysis of the processes identifies two technical complex objects (Figure 1):
- the request for quotation,
- the estimate

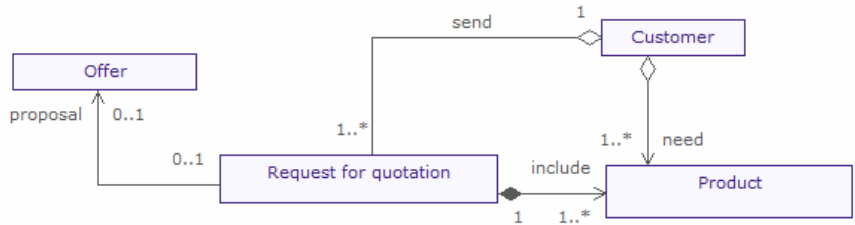

Figure 1: Extract of Class Diagram

These two items are liked: an estimate can not exist without a request for Quotation. In addition, each object has its proper information. The classes "estimate" and "request for Quotation" have common information. This information helps to link the two entities.

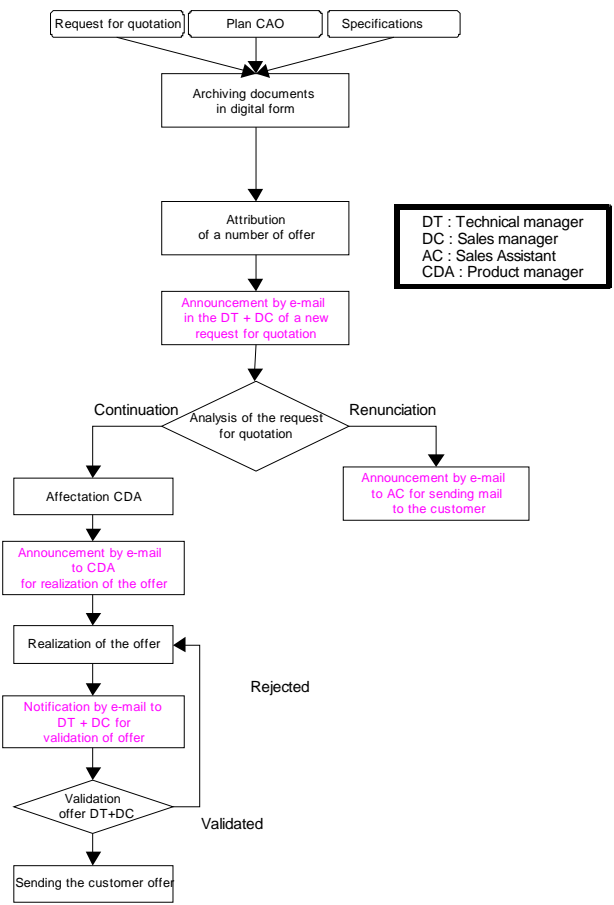

Figure 2: Process for supply management

Once the technical class identified, it's necessary to define its lifecycle. Every entity possesses a lifecycle. This lifecycle defines the statutes by which will move objects during different stages of the process. The following diagrams (Figure 3, Figure 4) showed sequences of statutes identified for each entity.

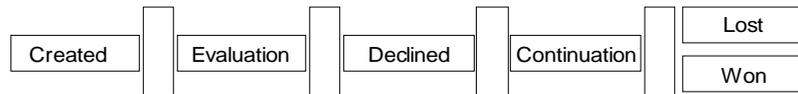

Figure 3: "Request For Quotation" lifecycle

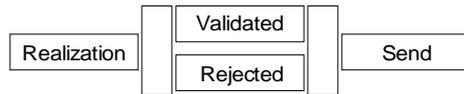

Figure 4: the lifecycle of the entity "Offer"

In a second time, it is necessary to identify those activities which result in change of state. For each class, the steps identified are:

"Request for Quotation" class:
— registration of the request for Quotation
— analysis of the request for quotation
— customer decision

"Offer" class:
— a project manager affectation
— Realization of the offer
— Validation of the offer
— Sending to customer the offer

In this step, we must define the experts who will validate the activities of this business processes. The UML diagram of use-case (Figure 5) allows us to identify the actors and the various activities on these classes. Also, we have classified these activities by defining a level of expertise. This approach allows for separate activities requiring specific skills. Thus, these activities (and actors associated with them) will be applied in the evaluation process.

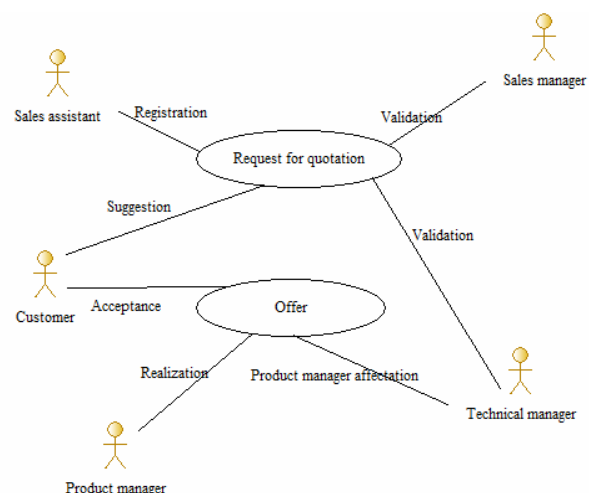

Figure 5: Use Case Diagram of the request for quotation

These steps have made explicit the business process of request for quotation in order to integrate it into a PLM system. Figure 6 presents a simplified extraction of the business process

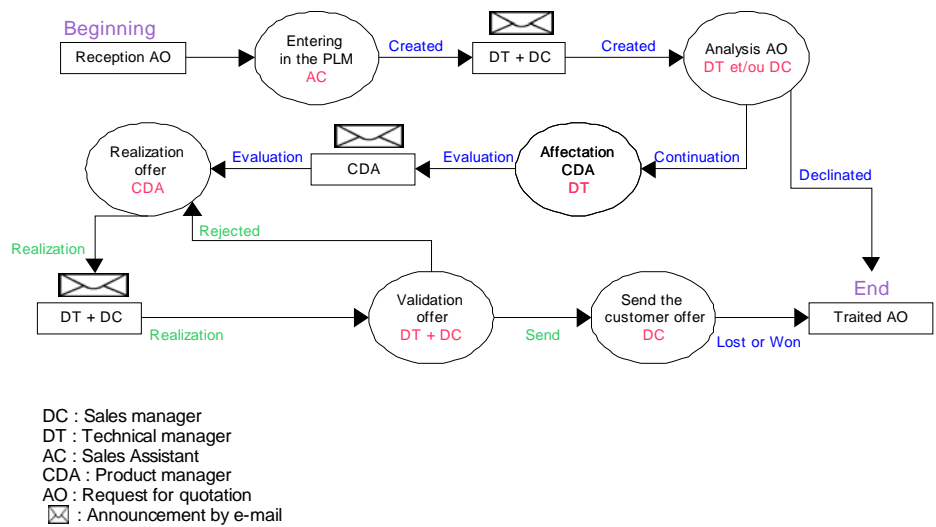

Figure 6: Simplified extraction of "Request For Quotation"

From the typology previously described, we can identify some criteria related to time, stability, genericity and measurability. The indicator construction is initiated by values provides by these characteristics (e.g. number of RFP declined, lost, won, average time between the change of state …)

At the same time, PLM systems have a capacity to manage traceability. From tracks generated on the activity, we developed an interface of indicators construction. It is standard indicators related to the objectives above. Besides the system suggests to build new indicators.

## 5. CONCLUSION

Our objective was to show how PLM systems allow an implementation of the business processes. Compared to a traditional approach process, it is necessary to consider the characteristics of the processes as well as the cycle. Our prospects are to constitute a whole of indicators associated with an objective. These indicators will allow a monitoring of the processes and a ROI calculation.

# LIST OF REFERENCES

*Book :*

Berrah, L. 2002. *L'indicateur de performance*, Cépaduès.

Debaecker D. 2004. *PLM La gestion collaborative du cycle de vie des produits*, Hermès science.

Imai, M. 1992. *Kaizen : La clé de la compétitivité japonnaise*, Eyrolles, 3$^{ème}$ ed..

Morley, C., Hugues, J., Leblanc, B., Hugues, O. 2005, *Processus métiers et S.I.. Evaluation, modélisation, mise en oeuvre*, Dunod.

Pernelle P, Lefebvre A. 2006. *Modélisation intégrée et pérennisation des connaissances dans une approche PLM, Ingénierie de la conception et cycle de vie des produits*, Hermès Science Publications.

*Conference paper :*

Berrah, L., Mauris G., Vernadat F. 2006. Industrial performance measurement: an approach based on the aggregation of unipolar or bipolar expressions, *International Journal of Production Research*, No. 18-19, p. 4145-4158.

Mhamedi, A.L., Lerch, C., Magano, H., Sonntag, M. 1995. Analyse des sytèmes de production par les activités : une approche pluridisciplinaire. Congrès international de Génie Industriel Montréal, Vol 1, p. 141-151.

Shelton, R.E. 1996, Business objects – workflow. Data Management Review.

Sudarsan, R., Fenves, S., Sriram, R. D., Wang, F. 2005. A Product Information Modeling Framework for Product Lifecycle Management, *Manufacturing Systems Integration Division, Manufacturing Engineering Laboratory*, National Institute of Standards and Technology, Gaithersburg, USA.

Mendoza, C., Delmond, M.H., Giraud, F., Löning, H. 2002, *Tableaux de bord et Balanced scorecards,* Groupe Revue Fidiciaire

*Journal article :*

Arkin, A. novembre 2002, Business Process Modeling Language, Version 1.0, http://bpmi.org.

OMG, 2006, '*BPMN 1.0' OMG Final Adopted Specification*, http://www.bpmn.org/

WFMC, 2005, 'XML Process Definition Language (XPDL)', *WFMC-TC-1025 FINAL,* http://www.wfmc.org/